\documentclass[preprint,nopreprintline,12pt]{elsarticle}
\usepackage{amsfonts}
\usepackage{amssymb}
\usepackage{mathrsfs}
\usepackage{upgreek}
\usepackage{graphicx,epstopdf} 
\usepackage[caption=false]{subfig}
\usepackage{pgfplots}
\usepackage{algorithmicx}
\usepackage{algpseudocode}
\usepackage{array} 
\usepackage{tabularx}
\usepackage{booktabs}
\usepackage{bm}
\usepackage{multirow}
\usepackage{multicol}
\usepackage[normalem]{ulem}
\usepackage{cancel}
\usepackage{braket}
\usepackage{amsmath}
\usepackage{cleveref}
\usepackage{algorithm}
\usepackage{bbm}

\newcommand{\rank}{\mathbbm{r}}
\newcommand{\dof}{\mathcal{M}}
\newcommand{\sample}{\mathcal{N}}
\newcommand{\OO}{\overline{\Osr}}
\newcommand{\loss}{\mathscr{L}}
\newcommand{\LL}{\overline{\mathcal{L}}}
\newcommand{\Osr}{\mathcal{O}}
\newcommand{\mS}{\mathcal{S}}
\newcommand{\grad}{g}
\newcommand{\Lbar}{\mathcal{L}}
\newcommand{\mT}{\mathcal{T}}

\newcommand{\Z}{\mathbb{Z}}

\newcommand{\R}{\mathbb{R}}
\newcommand\ham{\mathcal{H}}

\newcommand\xx{\pmb{x}}
\newcommand{\RR}{\pmb{R}}
\newcommand{\rr}{\pmb{r}}
\renewcommand{\l}{\ell}

\newcommand{\A}{\mathscr{A}}

\newcommand{\ind}{\nu}
\newcommand{\aceind}{\pmb{\nu}}

\newcommand{\Pa}{\uppi}

\newcommand\dd{~{\rm d}}
\newcommand\El{E_{\rm L}}
\newcommand\EE{\mathbb{E}}

\newcommand{\basis}{\phi}

\newcommand{\ace}{\pmb{A}}

\newcommand{\aceB}{\mathcal{B}}
\newcommand{\aceI}{\mathcal{I}}
\newcommand{\param}{\pmb{\theta}}

\newcommand{\wfbf}{\Psi_{\rm BF}}
\newcommand{\sse}{\mathbb{S}}

\newcommand{\trans}{\top}
\newtheorem{remark}{Remark}

\journal{J. Comput. Phys.}
\begin{document}
\begin{frontmatter}

\title{Stochastic Reconfiguration with Warm-Started SVD}


\author[inst1]{Dexuan Zhou}
\author[inst1]{Huajie Chen}

\affiliation[inst1]{organization={School of Mathematical Sciences, Beijing Normal University},
city = {Beijing},
postcode = {100875}, 
country = {China}
}

\author[inst2]{Cheuk Hin Ho}
\author[inst3,inst4]{Xin Liu}
\author[inst2]{Christoph Ortner}

\affiliation[inst2]{organization={Department of Mathematics, University of British Columbia},
city = {Vancouver},
postcode = {V6T 1Z2}, 
state = {BC},
country = {Canada}
}

\affiliation[inst3]{
  organization={State Key Laboratory of Mathematical Sciences, Academy of Mathematics and Systems Science, Chinese Academy of Sciences},
  city = {Beijing},
  postcode = {100190},
  country = {China}
}

\affiliation[inst4]{
  organization={School of Mathematical Sciences, 
                University of Chinese Academy of Sciences},
  city = {Beijing},
  postcode = {101408},
  country = {China}
}
\begin{abstract}
The combination of the variational Monte Carlo (VMC) method with deep learning wave function architectures has led to several successes in ground-state calculations of quantum many-body systems in recent years. 
However, commonly used stochastic gradient–based methods often perform poorly on these parameter training problems and typically lack convergence guarantees.
The stochastic reconfiguration (SR) method provides a robust pre-conditioner of the stochastic gradient, whose computational cost becomes prohibitive for large parameter spaces owing to the repeated inversion of large covariance matrices. 
To overcome this bottleneck, we propose a warm-started stochastic reconfiguration (WSSR) method, which integrates warm-start techniques from singular value decomposition (SVD) to refine low-rank approximations of the preconditioning matrix iteratively. 
Numerical experiments on typical atomic and molecular systems highlight the effectiveness of the WSSR method within VMC calculations.
\end{abstract}

\end{frontmatter}

\section{Introduction}
\label{sec:introduction}

Calculating the ground state of the many-electron Schr\"odinger equation remains an extremely challenging task. 
Traditional variational methods, such as Hartree–Fock \cite{lykos1963discussion}, configuration interaction \cite{shavitt1977method}, and multi-configuration self-consistent field \cite{hinze1973mc}, evaluate the electron energy and its first order variation by pre-computed one- and two-electron integrals.
In contrast, the variational Monte Carlo (VMC) method estimates the energy through stochastic methods and offers greater flexibility in the choice of wave function ansatz, such as the Slater-Jastrow and the Slater-Jastrow-Backflow models \cite{becca2017qmc}.
Recent advancements in machine learning-based wave function representations \cite{choo2020fermionic, giuse16, hermann2020dnn, pfau2020Ferminet, zhou2024multilevel} have significantly expanded the scope of wave function modeling and have gained increasing attention for their ability to capture complex correlations with both high flexibility and accuracy.
%
%
However, finding parameters that minimize the energy on a complex landscape remains a significant challenge. 
First-order optimization techniques, such as stochastic gradient descent \cite{amari1993backpropagation, ketkar2017stochastic} and Adam \cite{kingma2014adam, loshchilov2017decoupled}, often suffer from convergence difficulties, either failing to converge or converging very slowly.

One of the most effective approaches to date is the stochastic reconfiguration (SR) method, initially introduced in \cite{sorella2001generalized} for lattice systems and later extended to small atoms and molecules \cite{casula2004correlated,casula2003geminal}. 
The SR method leverages the geometric structure of the parameter space, updating parameters along the descent direction within the Fubini-Study metric \cite{becca2017qmc,goldshlager2024kaczmarz}.
Updating the parameters in the SR method requires inverting a covariance matrix $\mS$, which poses several key challenges in optimization \cite{lange2024architectures}:
(i) sampling introduces stochastic errors in estimating $\mS$, necessitating a large number of samples;
(ii) $\mS$ is often singular in practice, leading to instability in the optimization process;
(iii) the size of $\mS$ grows with the number of model parameters, making its inversion computationally expensive for large models.

To tackle these challenges, several modifications have been proposed. 
In \cite{martens2015optimizing}, an exponentially decaying averaging scheme is developed to estimate the $\mS$-matrix, incorporating historical estimates to improve stability. 
In \cite{chen2024empowering, hofmann2022role, park2020geometry, rende2024simple, schmitt2020quantum, sorella2007weak}, various regularization techniques are proposed to stabilize the optimization process, such as adding a diagonal shift to the $\mS$-matrix and setting a cut-off tolerance for small singular values.
Recent advancements have suggested modifications to the SR update rule, replacing the inversion of the $\mS$-matrix with the inversion of a smaller $\mT$-matrix, which reduces computational costs while maintaining accuracy \cite{chen2024empowering, goldshlager2024kaczmarz, rende2024simple}. 
Despite these improvements, a method that fully addresses these challenges (i.e. ensuring stability, enhancing efficiency, and enabling effective averaging schemes for the $\mS$-matrix) remains an open problem.

In this work, we propose a warm-started stochastic reconfiguration (WSSR) optimizer that integrates a warm-started singular value decomposition (SVD) algorithm into the SR method, to achieve efficient matrix inversion while maintaining the simplicity of averaging and regularization.
We observe that calculating the largest $k$ singular values can be formulated as an optimization problem \cite{liu2013limited}. 
This formulation allows for the application of a warm-start technique, which significantly reduces computational costs by leveraging the small changes to the target matrices in consecutive VMC iterations.
Instead of recomputing the entire SVD from scratch, one adjusts the existing decomposition from previous VMC steps, reusing the singular vectors to accelerate the process. 
This approach is particularly beneficial in our VMC framework as the $\mS$-matrices are frequently updated.
We further introduce a dynamic selection mechanism for the truncation rank to enhance optimization efficiency.

To demonstrate the effectiveness of the WSSR optimizer, we apply it to optimize an ACE wave function \cite{zhou2024multilevel} for both small atoms and molecule, a wave function architecture on which standard stochastic gradient methods perform particularly poorly. 
Our results show that WSSR can achieve comparable convergence as the state-of-art SR method, SPRING \cite{goldshlager2024kaczmarz} (see Section \ref{sr:variants}), while requiring less computational cost for each update step.

\vskip 0.2cm

{\bf Outline.}
The rest of this paper is organized as follows.
In Section \ref{sec:schrodinger}, we provide a brief review of the many-electron Schr\"odinger equation and the VMC method.
In Section \ref{sec:sr}, we present the standard SR method, as well as its various modifications.
In Section \ref{sec:SR:SVD}, we propose the WSSR optimizer and design an adaptive rank selection scheme.
In Section \ref{sec:numerics}, we show numerical simulations on typical atomic and molecular systems.
Finally, concluding remarks are given in Section \ref{sec:conclusions}.

\section{Background}
\label{sec:schrodinger}

\subsection{Many-electron Schr\"{o}dinger equation}
\label{sec:many-electron}
We consider a many-particle system consisting of $M$ nuclei and $N$ electrons. 
Let
$\RR := (\RR_1,\cdots,\RR_M)\in (\R^3)^M$ and $(Z_1,\cdots,Z_M)\in (\Z_+)^M$ 
denote positions and charges of nuclei, respectively. 
Meanwhile, let
$\rr := (\rr_1,\cdots,\rr_N)\in(\R^3)^N$ and $\pmb{\sigma} := (\sigma_1,\cdots,\sigma_N)\in(\Z_2)^N$
stand for spatial and spin coordinates of electrons. Here and throughout, we use 
$\xx := (\xx_1,\cdots,\xx_N)$ 
to denote the $N$-electron configuration with 
$\xx_i = (\rr_i, \sigma_i) \in \R^3 \times \Z_2~(i=1,\cdots,N)$.
Using the Born-Oppenheimer approximation \cite{born1927born}, the electron Hamiltonian of the system is given in atomic units by
\begin{equation}
\label{ham}
\ham := - \frac{1}{2} \sum_{i=1}^N \nabla_{\rr_i}^2 - \sum_{I=1}^M \sum_{i=1}^N \frac{Z_I}{|\rr_i - \RR_I|} + \sum_{1\leq i<j\leq N} \frac{1}{|\rr_i - \rr_j|} .
\end{equation}
The ground state of this system can be obtained by minimizing the energy functional 
\begin{eqnarray}
\label{gs:min}
E_0 = \min_{\Psi\in\A} E(\Psi) 
\quad{\rm with} \quad
E(\Psi) :=
\frac{\displaystyle\int_{\xx\in \big(\R^3\times\Z_2\big)^N} \Psi(\xx)\big(\ham\Psi\big)(\xx)\dd\xx}{\displaystyle\int_{\xx\in \big(\R^3\times\Z_2\big)^N}|\Psi(\xx)|^2\dd\xx} 
\end{eqnarray}
in the admissible class of the wave function
\begin{eqnarray}
\label{psi:class}
\A:= \Big\{ \Psi\in H^1((\R^3 \times \Z_2)^N):  \Psi~{\rm is~antisymmetric} \Big\} , 
\end{eqnarray}
and the integral with respect to the electron configuration $\xx$ means
\begin{equation*}
  \displaystyle\int_{\xx\in     \big(\R^3\times\Z_2\big)^N} 
    := \sum_{\sigma_1\in\Z_2}
    \cdots
    \sum_{\sigma_N\in\Z_2} 
    \int_{\R^3}\dd r_1
    \cdots
    \int_{\R^3} \dd r_N,   
\end{equation*}
and the anti-symmetry condition is given by
\begin{eqnarray}
\label{antisymmetric}
    \Psi \big( \xx_{\Pa(1)}, \cdots, \xx_{\Pa(N)} \big) 
    = (-1)^{\epsilon_\Pa} 
    \Psi \big( \xx_1, \cdots, \xx_N \big)
    \qquad \forall~\Pa\in S_N,
\end{eqnarray}
where $S_N$ denotes the permutation group and $\epsilon_\Pa$ denotes the parity of the permutation $\Pa\in S_N$.
Note that the ground-state energy $E_0$ is the lowest eigenvalue of the Hamiltonian $\ham$ with the corresponding eigenfunction belonging to $\A$.

In practical calculations, the wave function is parameterized using a physics-informed or machine-learning ansatz, $\Psi_{\param}\in\A$, with parameters $\param\in\R^{\dof}$.
Then the ground-state solution $E_0$ can be approximated by solving the following optimization problem,  
\begin{align}
\label{E:parm:L}
    \min_{\param\in\R^\dof} \loss(\param)
    \quad{\rm with}\quad \loss(\param) 
    := E(\Psi_{\param}), 
\end{align} 
which reduces the variational space $\A$ in \eqref{gs:min} into a lower-dimensional parameter space $\R^{\dof}$.
Nevertheless, we note that calculating $\loss(\param)$ still involves a high-dimensional integral.

\subsection{Variational Monte Carlo}
\label{sec:vmc}
Within the VMC framework, the loss function $\loss(\param)$ can be expressed as the expectation value of the ``local energy" as \cite{becca2017qmc} 
\begin{align}
\label{loss:exp}
    \loss(\param) = \displaystyle\int_{\xx\in \big(\R^3\times\Z_2\big)^N} \El(\xx;\Psi_{\param}) \cdot P(\xx;\Psi_{\param}) \dd\xx
    = \EE_{\xx\sim P(\cdot;\Psi_{\param})} \big[ \El(\xx;\Psi_{\param}) \big] ,
\end{align}
where the probability density $P(\xx;\Psi)$ and the local energy $\El(\xx;\Psi)$ are, respectively, defined as
\begin{align}
\label{vmc:P}
    P(\xx;\Psi) 
    := 
    \frac{\big|\Psi(\xx)\big|^2}{
    \displaystyle
    \int_{\xx\in \big(\R^3\times\Z_2\big)^N}
    \big|\Psi(\xx)\big|^2\dd\xx}
    \quad{\rm and}\quad
    \El(\xx;\Psi) 
    := \frac{\big(\ham\Psi\big)(\xx)}{\Psi(\xx)}. 
\end{align}
To optimize the loss function, the gradient of $\loss(\param)$ is also required. 
A similar derivation \cite{lin2021explicitly} yields 
\begin{align}
\label{gradient:exp}
    \grad(\param) 
    := 
    \nabla_{\param}\loss(\param)
    = 
    2~\EE_{\xx\sim P(\cdot;\Psi_{\param})} 
    \Big[ 
    \big(
    \nabla_{\param}
    \log|\Psi_{\param}(\xx)|
    \big) \big(\El(\xx;\Psi_{\param}) 
    - \loss(\param)\big) 
    \Big] .
\end{align}
In practice, both $\loss(\param)$ and $g(\param)$ are estimated stochastically using Markov-Chain Monte Carlo sampling \cite{foulkes2001qmc} to generate $\sample$ samples from $P(\xx;\Psi)$, denoted as $\sse_{\sample} := \{\xx_1, \cdots, \xx_\sample\}$.
Unbiased estimates for the loss $\loss(\param)$ and the gradient $g(\param)$ are then given by 
\begin{align}
\label{loss:vmc}
    &
    \loss(\param) 
    \approx 
    \frac{1}{\sample}\sum_{\xx\in\sse_\sample} \El(\xx;\Psi_{\param}) 
    =: 
    \loss_\sample(\param),
    \\[1ex]
\label{gradient:vmc}
    & \grad(\param) 
    \approx 
    \frac{2}{\sample}\sum_{\xx\in\sse_\sample} \Big( \partial_{\param}\log|\Psi_{\param}(\xx)| \Big) \Big( \El(\xx;\Psi_{\param}) 
    - \loss_\sample(\param) \Big)
    =: 
    \grad_\sample(\param).
\end{align}
Note that $\grad_\sample(\param)$ is an unbiased estimator of the gradient $\nabla_{\param}\loss(\param)$, which however is not equal to the gradient of $\loss_\sample(\param)$.

Now $\loss(\param)$ can be minimized using optimization methods, such as stochastic gradient descent (SGD) \cite{amari1993backpropagation, ketkar2017stochastic}, i.e., 
the parameter $\param^{(k)}$ at the $k$-th iteration is updated as
\begin{align} 
\label{sgd} \param^{(k+1)} = \param^{(k)} - \eta^{(k)} \cdot \grad_\sample(\param^{(k)}), \end{align}
where $\eta^{(k)} > 0$ is the learning rate.
However, SGD-based methods, including modifications like Adam \cite{kingma2014adam, loshchilov2017decoupled}, generally perform poorly for complex systems and complex wave function architectures, as the updates often get stuck,  oscillating back and forth along steep regions of the energy landscape instead of converging along the flatter directions \cite{park2020geometry}. 
We have observed particularly severe effects of this kind for the ACE wave functions introduced in \cite{zhou2024multilevel}. (We speculate that this is due to the hierarchical features that ACE employs.)
In practice, the Stochastic Reconfiguration (SR) method \cite{casula2004correlated, casula2003geminal, sorella2001generalized} is frequently used, as it takes advantage of the geometry of the parameterization space, offering a more reliable and efficient optimization scheme.

\begin{remark}[Spin Models]
\label{remark:spin}
Beside the many-electron Schr\"odinger equation, spin models have also attracted growing attention within the VMC framework, particularly with the integration of machine learning architectures \cite{chen2018equivalence, kochkov2021learning, liang2018solving}. 
All algorithms proposed in this work can be readily extended to spin models.
\end{remark}

\section{Stochastic Reconfiguration}
\label{sec:sr}

\subsection{Standard Stochastic Reconfiguration method}
\label{sec:standardsr}
The SR method is based on the idea of performing an imaginary time evolution and finding parameter updates by minimizing the Fubini-Study distance at each small imaginary time step \cite{becca2017qmc, goldshlager2024kaczmarz}. This results in a preconditioned gradient $\mS\big(\param\big)^{-1} \grad\big(\param\big)$, where the preconditioner $\mS(\param)$ is represented as a covariance matrix, 
\begin{equation}
\label{eq:fim}
    \mS(\param) 
    := 
    \EE_{\xx\sim P(\cdot;\Psi_{\param})} \left[
    \Osr(\xx; \param) \cdot \Osr(\xx; \param)^{\trans}
    \right]  
    \in \R^{\dof\times \dof}, 
\end{equation}
with
\begin{eqnarray}
\label{Osr}
    \Osr(\xx;\param) = \left(
    \partial_{\param}   
    \log    
    \big|
    \Psi_{\param}(\xx)|  
    - 
    \EE_{\xx\sim P(\cdot;\Psi_{\param})} 
    \Big[ 
    \partial_{\param}\log
    \big|\Psi_{\param}(\xx)| 
    \Big]
    \right)\in \R^{\dof}.
\end{eqnarray}
In practice, the expectation in \eqref{eq:fim} can be stochastically estimated as follows
\begin{eqnarray}
\label{eq:SN}
    \mS(\param)     \approx \sum_{\xx\in\sse_\sample}   
    \Osr(\xx;\param) \cdot \Osr(\xx;\param)^{\trans} 
    =: \mS_\sample(\param) .
\end{eqnarray}
Consequently, rather than using \eqref{sgd}, the parameter update is expressed in the form
\begin{eqnarray}
\label{eq:param:update2}
    \param^{(k+1)}  
    = 
    \param^{(k)}  
    - \eta^{(k)} \mS_{\sample}
    \big(
    \param^{(k)}
    \big)^{-1} \grad_{\sample}
    \big(
    \param^{(k)}
    \big) .
\end{eqnarray}

\begin{remark}
[More interpretation of the $\mS$-Matrix]
\label{remark:S-matrix}
The $\mS$-matrix in \eqref{eq:fim} can be understood from several perspectives. 
By definition, the $\mS$-matrix represents the covariance matrix of the gradient of the logarithm of the wave function, quantifying the variance of gradient estimates across different samples. 
Moreover, the SR method, often referred to as the natural gradient method, links the $\mS$-matrix to the Fisher Information Matrix (FIM) \cite{pfau2020Ferminet, lin2021explicitly}, encodes the geometric structure of the parameter space, allowing the gradient directions to be adjusted for more efficient parameter updates. 
\end{remark}

In practice, one can express $\mS_{\sample}(\param)$ and $\grad_\sample(\param)$ in an alternative matrix form.
If we define   
\begin{align}
\label{eq:Osrn}
    \Osr(\sse_{\sample};\param) &= \frac{1}{\sqrt{\sample}} 
    \begin{bmatrix}
        \Osr(\xx_1;\param) & \cdots & \Osr(\xx_\sample;\param)
    \end{bmatrix} \in \R^{\dof \times \sample}
    \quad{\rm and}
    \\
\label{eq:Lbarn}
    \Lbar(\sse_{\sample};\param) &= \frac{1}{\sqrt{\sample}} 
    \begin{bmatrix}
        \Lbar(\xx_1;\param) & \cdots & \Lbar(\xx_\sample;\param)
    \end{bmatrix} \in \R^{\sample}, 
\end{align} 
where $\Lbar(\xx; \param)$ is given by $\Lbar(\xx; \param) := \El(\xx;\Psi_{\param}) - \loss_\sample(\param)$, then $\mS_{\sample}(\param)$ and $\grad_\sample(\param)$ can then be written as
\begin{align*}
    \mS_{\sample}(\param) = \Osr(\sse_{\sample};\param) \Osr(\sse_{\sample};\param)^{\trans}
    \qquad{\rm and}\qquad
    \grad_\sample(\param) = 2\Osr(\sse_{\sample};\param) \Lbar(\sse_{\sample};\param). 
\end{align*}
These representations will be used in the following.

\subsection{Variants of the SR method}
\label{sr:variants}

The SR update \eqref{eq:param:update2} involves an inversion of the $\mS$-matrix, which comes with three key issues (as discussed in Introduction): stochastic error, invertibility, and high computational cost.
To address these issues, several modifications have been proposed.


For the first issue related to statistical errors, a common approach is to use averaging techniques that incorporate information from previous iterations \cite{martens2015optimizing,pfau2020Ferminet}, i.e., 
$\mS_{\sample}$ in \eqref{eq:param:update2} is replaced by $\mS^{\rm ave}_{\sample}$ of the following form 
\begin{eqnarray}
\label{eq:ave_s}
\mS^{\rm ave}_{\sample}\big(\param^{(k)}\big) = 
\begin{cases}
        \mS_\sample\big(\param^{(k)}\big), & \mathrm{if}~k = 1 ,
        \\[1ex]
        \delta \mS^{\rm ave}_{\sample}\big(\param^{(k-1)}\big) + (1-\delta) \mS_\sample\big(\param^{(k)}\big),& \mathrm{otherwise},
\end{cases} 
\end{eqnarray}
where $\delta\in(0,1)$ represents the weight of averaging. 


For the second issue on invertibility of $\mS$, a broad class of regularization schemes have been developed \cite{park2020geometry, rende2024simple}, which replace $\mS_{\sample}$ in \eqref{eq:param:update2} by $\mS_{\sample}^{\rm reg}$ of the form
\begin{eqnarray*}
    \mS^{\rm reg}_{\sample}(\param) 
    = \mS_{\sample}(\param) + \epsilon I
    \quad{\rm or}   \quad
    \big(\mS^{\rm reg}_{\sample}(\param)\big)_{ij} = 
    \begin{cases}
    (1+\epsilon)\big(\mS_{\sample}(\param)\big)_{ii}, &\mathrm{if}~ i = j ,
    \\[1ex]
    \big(\mS_{\sample}(\param)\big)_{ij}, & \mathrm{otherwise},
    \end{cases}
\end{eqnarray*}
where $\varepsilon > 0$ is a hyper-parameter for regularization.
A different type of regularization replaces $\mS_{\sample}^{-1}$ with its pseudo-inverse, $\mS_{\sample}^\dagger$ \cite{chen2024empowering, hofmann2022role}, i.e., 
\begin{eqnarray}
\label{eq:svd0}
    \mS_{\sample}(\param)^\dagger 
    = 
    U(\param) \Sigma(\param)^\dagger U(\param)^{\trans} ,
\end{eqnarray}
where 
$\mS_{\sample}(\param) = U (\param)\Sigma(\param) U(\param)^{\trans}$ 
is the SVD calculation of $\mS_{\sample}(\param)$ 
and 
$\Sigma(\param)^\dagger$ is a cutoff of $\Sigma(\param)^{-1}$ with diagonal elements
\begin{eqnarray}
\label{eq:sigmaplus}
    \sigma_i^\dagger = 
    \begin{cases}
    1/\sigma_i, & \mathrm{if}~|\sigma_i|\geq \varepsilon_{\rm tol} |\sigma_{\max}| , 
    \\[1ex]
    0, & \mathrm{otherwise}.
    \end{cases} 
\end{eqnarray}
Here, $\sigma_i$ are the diagonal elements of $\Sigma(\param)$, $\sigma_{\max}$ is the maximum singular value, and $\varepsilon_{\rm tol}>0$ is the relative tolerance used to truncate small singular values.
We emphasize that the choice of regularization schemes has a significant impact on convergence: over-regularization leads to underfitting and inefficiency \cite{lange2024neural}, while under-regularization may cause instability.

To address the third issue, the computational cost of inverting $\mS$, several variants of the SR method have been developed, including Krylov-type methods \cite{giuse16, neuscamman2012optimizing} and the Kronecker-Factored Approximate Curvature (K-FAC) method \cite{pfau2020Ferminet, martens2015optimizing}. 
Additionally, the recently developed minimum-step Stochastic Reconfiguration (MinSR) method \cite{chen2024empowering, chen2023efficient, rende2024simple} exploits the relation
\begin{eqnarray} 
\label{eq:mT}
\mS_{\sample}(\param)^{-1} \grad_{\sample}(\param) = 2\Osr(\sse_{\sample};\param) \mT_{\sample}(\param)^{-1} \Lbar(\sse_{\sample};\param), 
\end{eqnarray}
where $\mT_{\sample}(\param) := \Osr(\sse_{\sample};\param)^{\trans} \Osr(\sse_{\sample};\param)$, and $\Osr(\sse_{\sample};\param)$, $\Lbar(\sse_{\sample};\param)$ are given in \eqref{eq:Osrn} and \eqref{eq:Lbarn}, 
respectively. 
This approach circumvents the need to invert
$\mS_{\sample}(\param)\in \R^{\dof \times \dof}$ by instead inverting the smaller matrix $\mT_{\sample}(\param) \in \R^{\sample \times \sample}$, which is computationally advantageous as $\dof \gg \sample$ in most cases.

As a substantial modification of MinSR, the Subsampled Projected-Increment Natural Gradient (SPRING) method \cite{goldshlager2024kaczmarz} introduces a momentum term to the parameter updating formula,
\begin{align} 
\label{eq:param:spring} 
\param^{(k+1)} - 
\param^{(k)} & = \phi(\param^{(k)}) + \mu \big(\param^{(k)} - \param^{(k-1)}\big), \quad \text{with}
\\[1ex] \nonumber 
\phi(\param^{(k)}) & =- \eta^{(k)} \Osr_{\sample}\big(\sse_{\sample};\param^{(k)}\big) \mT_{\sample}(\param)^{-1} \tilde{\Lbar}(\sse_{\sample};\param^{(k)}), \quad \text{where}
\\[1ex] \nonumber 
\tilde{\Lbar}(\sse_{\sample};\param^{(k)}) &:= \Lbar(\sse_{\sample};\param^{(k)}) - \mu \Osr_{\sample}\big(\sse_{\sample};\param^{(k)}\big)^{\trans} (\param^{(k)}-\param^{(k-1)}). 
\end{align} 
This removes the momentum component that conflicts with the current subproblem, controlled by a decay parameter $\mu \in [0,1)$. When $\mu = 0$, the method reduces to MinSR. 
MinSR updates parameters based on single minibatch, whereas SPRING exploits the history of past optimization. It is reasonable to doubt that averaging past gradients, instead of averaging the S-matrix, may successfully mitigate the stochastic error.

The goal of this work is to develop a method that simultaneously addresses all three issues within a unified framework, and to provide a comparison with existing approaches.

\section{Stochastic reconfiguration with warm-started SVD}
\label{sec:SR:SVD}

In this section, we propose a Warm-Started Stochastic Reconfiguration (WSSR) method, whose main idea is to utilize historical information to accelerate the convergence of the iterative SVD algorithm at each optimization step. 
In what follows, we first present a framework that provides a low-rank approximation of the ``averaged" covariant matrix \eqref{eq:ave_s} in Section~\ref{sec:idea}, then combine the warm-started SVD algorithm to solve the bottleneck in computing the low-rank approximation in Section~\ref{sec:warmSVD}. To further enhance efficiency and automate hyper-parameter tuning, we introduce an adaptive rank selection scheme in Section~\ref{sec:rank}.


\subsection{Low-rank approximation of the averaged $\mS$-matrix}
\label{sec:idea}

In this subsection, we develop a low-rank approximation method for the averaged $\mS$-matrix, which can be viewed as an efficient combination of \eqref{eq:ave_s} and \eqref{eq:svd0}. 

Let $\rank \in \mathbb{Z}_+$ be a chosen rank.
At the $k$-th VMC step, rather than directly storing and averaging the gradient and $\mS$-matrix, we use a matrix $\OO^{(k)} \in \mathbb{R}^{\dof \times \rank}$ and a vector $\LL^{(k)} \in \mathbb{R}^{\rank}$ to retain the essential information from the relevant historical data.

For $k=0$, we set $\OO^{(0)}\in\R^{\dof \times \rank}$ by a zero matrix and $\LL^{(0)}\in\R^{\rank}$ by a zero vector, respectively.
They are updated in the VMC iteration as follows.
At the $k$-th step ($k > 0$), after evaluating $\Osr(\sse_{\sample}; \param^{(k)})$ and $\Lbar(\sse_{\sample}; \param^{(k)})$ defined in \eqref{eq:Osrn} and \eqref{eq:Lbarn}, we set
\begin{align}
\label{eq:Ohat}
    \hat{\Osr}^{(k)} 
    &:=  \left[ 
    \sqrt{\delta} ~ \OO^{(k-1)} , ~ \sqrt{1-\delta} ~ \Osr(\sse_{\sample}; \param^{(k)}) \right] 
    \in \R^{\dof \times (\rank + \sample)}, 
    \\[1ex]
\label{eq:Lhat}
    \hat{\Lbar}^{(k)} 
    &:= \left[ 
    \sqrt{\delta} ~ \LL^{(k-1)}, ~ \sqrt{1-\delta} ~ \Lbar(\sse_{\sample}; \param^{(k)}) 
    \right] \in \R^{\rank + \sample} ,
\end{align}
where $\delta \in (0,1)$ acts as a averaging weight.
Then we {calculate the rank-$\rank$ truncated SVD for $\hat{\Osr}^{(k)}$ as 
\begin{align}
\label{svd:det}
\hat{\Osr}^{(k)} \approx U^{(k)} \Sigma^{(k)} V^{(k)}, 
\end{align}
where $U^{(k)} \in \R^{\dof \times \rank}$, $V^{(k)} \in \R^{\rank \times (\rank + \sample)}$ are orthogonal matrices, and $\Sigma^{(k)}$ is a diagonal matrix containing the largest $\rank$ singular values $s_i^{(k)}$ in descending order, i.e., 
$s_1^{(k)} \geq s_2^{(k)} \geq \cdots \geq s_{\rank}^{(k)} > 0$.
Finally we can construct the new $\OO^{(k)}$ and $\LL^{(k)}$ by
\begin{align}
\label{eq:Obar}
\OO^{(k)} &:= U^{(k)} \Sigma^{(k)} \in \R^{\dof \times \rank}, 
\\[1ex]
\label{eq:Lbar}
\LL^{(k)} &:= V^{(k)} \hat{\Lbar}^{(k)} \in \R^{\rank} .
\end{align}

With the above computations, we are able to replace the gradient $\grad_{\sample}(\param^{(k)})$ and covariance matrix $\mS_{\sample}(\param^{(k)})$ with the following $\overline{\grad}^{(k)}$ and $\overline{\mS}^{(k)}$,
\begin{align}
 \label{eq:barg}
    \overline{\grad}^{(k)} 
    &:= 
    \hat{\Osr}^{(k)}
    \hat{\Lbar}^{(k)} \in \R^{\dof},
    \\[1ex]
    \label{eq:barS}
\overline{\mS}^{(k)} 
    &:= 
    \OO^{(k)}
    \big(\OO^{(k)}\big)^{\trans}
    \in \R^{\dof \times \dof} ,
\end{align}
which approximate the averaged gradient vector and the averaged $\mS$-matrix, respectively.
These approximations enable efficient computation and storage of the gradient vectors and covariance matrices, by capturing the essential information while reducing computational complexity and memory usage.
\begin{remark}
[Connection to the averaging scheme]
\label{remark:averageS}
If we perform an exact SVD at each step, with no truncation applied, the update rule for the $\mS$-matrix, as shown in \eqref{eq:barS}, becomes equivalent to
\begin{align*}
    \overline{\mS}^{(k)} 
    = \delta ~ \overline{\mS}^{(k-1)} 
    + (1-\delta) ~ \mS_{\sample}(\param^k),
\end{align*}
which aligns with the averaging scheme in \eqref{eq:ave_s}.
Similarly, the gradient $\overline{\grad}^{(k)}$ is updated in an averaging manner, 
\begin{align*}
    \overline{\grad}^{(k)} = \delta ~ \overline{\grad}^{(k-1)} + (1-\delta) ~ \grad_{\sample}(\param^k). 
\end{align*}
However, in practical calculations it is not feasible to keep increasing the rank in this manner. Therefore, we introduce a rank selection strategy later in \cref{sec:rank} to reduce computational costs.
\end{remark}

In practical calculations, we typically have $\rank < \min(\dof, \sample)$, which means that $\overline{\mS}^{(k)}$ is singular. 
As a result, additional regularization schemes are necessary to avoid the singularity issue and ensure stable parameter updates.
Note that only the first $\rank$ singular values of $\overline{\mS}^{(k)}$, denoted $\sigma^{(k)}$, are obtained from the truncated SVD calculation of $\hat{\Osr}^{(k)}$. Specifically, $\sigma_i^{(k)} = (s_i^{(k)})^2$ for $1\leq i\leq\rank$, where $s_i^{(k)}$ is the singular value of $\hat{\Osr}^{(k)}$. 
Therefore, we truncate the remaining singular values of the $\mS$-matrix to be $\underline{\sigma}$, where $\underline{\sigma}$ is a hyperparameter for regularization.

The low-rank approximation is now written as
\begin{align}
\label{eq:updated_sr_1}
    \param^{(k+1)}  
    & = 
    \param^{(k)}  
    - \eta^{(k)} \big( 
    \overline{\mS}^{(k) \dagger} + 
    \underline{\sigma}^{-1}
    \overline{\mS}^{(k)\perp} 
    \big) 
    \overline{\grad}^{(k)} , 
\end{align}
where $\eta^{(k)} > 0$ is the learning rate, $\overline{\mS}^{(k) +}$ and $\overline{\mS}^{(k) \perp}$ are given by
\begin{eqnarray*}
    \overline{\mS}^{(k) \dagger}
    :=U^{(k)} 
    \big(\Sigma^{(k)}\big)^{-2} \big(U^{(k)}\big)^{\trans} 
    \quad{\rm and} \quad
    \overline{\mS}^{(k) \perp}
    :=I-U^{(k)}   
    \big(U^{(k)}\big)^{\trans} .
\end{eqnarray*}
The update scheme presented in \eqref{eq:updated_sr_1} incorporates a low-rank approximation, weighted averaging, and regularization techniques to enhance computational efficiency, accuracy, and stability of the optimization process. This combination allows for more efficient handling of large-scale optimization problems. The main computational overhead of this approach arises from the partial SVD calculation in \eqref{svd:det}, which has a complexity of $O(\dof \sample\rank )$. 

\begin{remark}
[SVD with Randomized Sketching]
\label{remark:randomizedsketch}
Randomized sketching \cite{halko2011finding,martinsson2020randomized} is a stochastic strategy to handle large-scale matrices computations, by projecting data into lower-dimensional spaces with random matrices.
Methods equipped with randomized sketching reduce the dimensionality of large matrices or datasets, improving efficiency for tasks such as solving linear systems, low-rank matrix approximations, and regression. 
This strategy reduces the computational complexity of traditional methods, such as the simple subspace iteration (SSI)  and so on, in calculating partial SVD to $O(\dof \sample \log\rank)$, which comes at the cost of potential accuracy loss.
In \ref{sec:SVD:random}, we present more details for performing partial SVD with randomized sketching. 
In the following, we refer to the SR method that computes the partial SVD \eqref{svd:det} with randomized sketching as the Randomized Sketching Stochastic Reconfiguration (RSSR) method.
\end{remark}

\subsection{Warm-started SVD}
\label{sec:warmSVD}

Building on the previous iteration, we observe that computation of the largest singular values can be updated via simple subspace iteration (SSI) or reformulated as an optimization problem \cite{bathe2013subspace, saad2016analysis}. 
By concurrently updating the singular vectors during the solution process, we implement a warm-start strategy that capitalizes on the incremental changes between consecutive iterations to substantially reduce computational costs.

The SSI method generalizes the power method for the case of $\rank=1$ to compute the first $\rank$ dominant eigenvalues of a matrix through iterative matrix multiplications and orthogonalization. 
Fundamentally, SSI is based on computing eigenpairs of the symmetric matrix $\hat{\Osr}^{(k+1)} \big( \hat{\Osr}^{(k+1)} \big)^\trans$, which can be formulated as maximizing the Rayleigh–Ritz function under orthogonality constraints, 
\begin{equation}
\label{max:SSI}
\max_{U \in \mathbb{R}^{\dof \times \rank}} \|\hat{\Osr}^{(k+1)} U\|_{\rm F} \quad \text{subject to} \quad U^\trans U = I.
\end{equation}
We consider an iterative algorithm by refining an initial subspace spanned by $U^{(k)}$, i.e., one starts with an initial subspace $U^{(k+1), 0} := U^{(k)} \in \mathbb{R}^{\dof \times \rank}$, and updates it iteratively by
\begin{equation*}
   U^{(k+1), i} = \operatorname{orth} \left( \hat{\Osr}^{(k+1)} \big( \hat{\Osr}^{(k+1)} \big)^\trans U^{(k+1), i-1} \right) 
   \qquad{\rm for}~ i=1,2,\cdots . 
\end{equation*}
Here, $\operatorname{orth}(V)$ denotes the set of orthonormal bases for the range space of $V$. 
The method ensures that, after a few iterations, $U^{(k+1),i}$ converges to the first $\rank$ dominant singular vectors of $\hat{\Osr}^{(k+1)} \big( \hat{\Osr}^{(k+1)} \big)^\trans$. 
Through straightforward calculations, we obtain the SVD of $\hat{\Osr}^{(k+1)}$, specifically determining $U^{(k+1)}, \Sigma^{(k+1)}$, and $V^{(k+1)}$ in \eqref{svd:det} for the $(k+1)$-th step.
Its detailed procedure is presented in Algorithm \ref{algorithm-ssi}.

\begin{algorithm}[htbp!]
\caption{SVD with Subspace Iteration (SSI-SVD)}
\label{algorithm-ssi}
{\bf Input:}
Matrix $\hat{\Osr}^{(k+1)} \in \mathbb{R}^{\dof \times (\sample+\rank)}$, rank $\rank$, maximum number of iterations $m$, Initial guess $U^{(k)}$
\begin{algorithmic}[1]
\State Let $U^{(k+1), 0}=U^{(k)}$
\For{$i = 1$ to $m$}
 \State Compute QR decomposition $ U^{(k+1), i} = Q^{(k+1), i} R^{(k+1), i}$
\State Compute $V^{(k+1), i} = (\hat{\Osr}^{(k+1)})^\trans Q^{(k+1), i}$
\State 
Compute $U^{(k+1), i} = \hat{\Osr}^{(k+1)} V^{(k+1), i}$
\EndFor

\State Perform the QR decomposition: $V^{(k+1), m} = Q R$  
\State Sort $\text{diag}(R)$ in descending order by absolute value:  
${\rm ind} = \operatorname{sort}(\text{diag}(R))$ 
\State Rearrange $U^{(k+1), m}$, $\text{diag}(R)$, and $Q$ according to $\text{ind}$ to obtain 
$U^{(k+1)}$, $\Sigma^{(k+1)}$, and $V^{(k+1)}$.
\end{algorithmic}
{\bf Output:} $U^{(k+1)}$, $\Sigma^{(k+1)}$, $V^{(k+1)}$ 
\end{algorithm}

For the complexity of the SSI-SVD algorithm, the cost of matrix-block multiplications in the subspace iteration is $\mathcal{O}(m\dof \sample \rank)$, and the cost of orthonormalization (i.e., by QR factorization) is $\mathcal{O}(m\dof \rank^2)$, where $m$ denotes the number of iteration steps. 
The computational efficiency largely depends on the number of subspace iteration steps, and hence the quality of the initial guess.

In our framework, when the averaging weight $\delta$ is large and the learning rate for parameter updates is small, the matrix $\hat{\Osr}^{(k+1)}$ generated by \eqref{eq:Ohat} is very close to $\hat{\Osr}^{(k)}$ from the previous step. Consequently, the singular vectors of $\hat{\Osr}^{(k+1)}$ and $\hat{\Osr}^{(k)}$ are likely to be very similar due to the perturbation theory of eigenvalue decomposition. 
To verify this, at each VMC step $k$ for the Helium atom, we computed a high-accuracy SVD of $\hat{\Osr}^{(k)}$ and stored its leading $\rank$ left singular vectors $U^{(k)}$. 
We then monitored $\|s^{(k+1)} - s^{(k)}\|$ (with $s^{(k)}$ containing the first $\rank$ singular values) and $\|P^{(k+1)} - P^{(k)}\|$ (with $P^{(k)} := U^{(k)} U^{(k)\trans}$ in Table~\ref{tab:sr-stability-he}.
We observe that during the optimization procedure of VMC, the differences in singular values remain at the $10^{-3}$ level, while the projector differences are typically between $0.01$ and $0.1$.
Thus by employing the warm-started strategy, the singular vectors $U^{(k)}$ from the previous step can serve as an excellent initialization for the subspace iteration on $\hat{\Osr}^{(k+1)}$, significantly reducing the computational cost. 

\begin{table}[htbp!]
\centering
\footnotesize
\begin{tabular}{c c c c c}
\toprule
& $4{,}000$ & $8{,}000$ & $16{,}000$ & $32{,}000$ \\
\midrule
$\big\|s^{(k+1)} - s^{(k)}\big\|$
& $4.06{\times}10^{-3}$
& $8.66{\times}10^{-3}$
& $4.55{\times}10^{-3}$
& $3.32{\times}10^{-3}$
\\[5pt]
$\big\|P^{(k+1)} - P^{(k)}\big\|$
& $0.027$
& $0.111$
& $0.017$
& $0.036$
\\
\bottomrule
\end{tabular}
\caption{Differences between consecutive SR matrices $\hat{\Osr}^{(k)}$
for the Helium example.}
\label{tab:sr-stability-he}
\end{table}

\begin{remark}
[Iterative Methods for Large Matrices]
\label{remark:iterate}
Several methods have been proposed to accelerate the convergence of the SSI method, several iterative algorithms such as Arnoldi \cite{lehoucq2001implicitly}, Lanczos \cite{sorensen1997implicitly}, and Jacobi–Davidson \cite{stathopoulos1994davidson}, which efficiently handle large sparse or structured matrices. 
Usually, these approaches do not benefit from a warm-start strategy. 
For large unstructured matrices, specialized methods like the limited memory block Krylov subspace optimization (LMSVD) \cite{liu2013limited} accelerate SVD computations by using a dynamically selected subspace from previous iterations.
\end{remark}

\subsection{Adaptive rank selection scheme}
\label{sec:rank}

The remaining issue is to select a rank truncation, i.e., the hyper-parameter $\rank$, for the SVD at each iteration step.
This plays a crucial role in balancing accuracy and computational efficiency: 
If the rank is set too high, the model retains unnecessary components, leading to excessive computational cost; 
Conversely, if the rank is set too low, essential information of the $\mS$-matrix may be lost, resulting in a poor preconditioner for the gradients.
To navigate this trade-off, we introduce an adaptive strategy.

Intuitively, we truncate the singular values based on their ratio to the maximum singular value, i.e., we set $\rank$ to be the cardinality of
\begin{equation*}
\mathcal{I}_{\rm reg} := \big\{i: r_{\rm reg} \leq \sigma_i/\sigma_{\max} \big\} , 
\end{equation*}
where the $\sigma_i$ are the singular values of $\mS$ in descending order, $\sigma_{\max} = \sigma_1$ is the largest singular value, and $r_{\rm reg} \in (0, 1)$ is the truncation ratio.

To further control the computational cost, we compute only the first $\rank_{\max}$ singular values. 
Specifically, the rank is determined by using the following criterion:
\begin{equation}
\label{eq:ruler}
\rank = \max\Big\{i\in \mathcal{I}_{\rm reg}: i \leq \rank_{\max}\Big\} . 
\end{equation}
The maximum rank $\rank_{\max}$ is adjusted dynamically by an adaptive algorithm. 
We begin with an initial maximum rank, $\rank_{\max} = \rank_0$. 
If a predefined criterion is met, say $\rank_{\max} \in \mathcal{I}_{\rm reg}$, the rank is expanded by $\rank_{\max} \rightarrow (1+\epsilon) \rank_{\max}$. 
Our numerical experiments show that the maximum rank usually will not explode, see Section \ref{sec:numerics}. 

The proposed WSSR optimizer with adaptive rank selection scheme is summarized in Algorithm \ref{algorithm:wssr}. 

\begin{algorithm}[h!]
\caption{~~Warm-Started Stochastic Reconfiguration (WSSR)}
\label{algorithm:wssr}
\vskip 0.1cm
{\bf Input:}
Maximum number of iterations $k_{\max}$, 
learning rate schedule $\eta^{(k)}$, 
batch size $\sample$, 
averaging weight $\delta$, 
regularization parameter $\underline{\sigma}$, 
relative cutoff $r_{\rm reg}$, 
initial maximum rank $\rank_0$, 
rank expansion factor $\epsilon$,
initial parameter $\param^{(0)}$
\begin{algorithmic}[1]
\State 
    $\rank_{\max} = \rank_0$
\For{$k = 1, \dots, k_{\max}$}
    \State 
        Sample $\sse_{\sample}$ from $P(\cdot; \Psi_{\param^{(k)}})$
    \State 
        Compute $\Osr(\sse_{\sample};\param)$ and $\Lbar(\sse_{\sample};\param)$ with \eqref{eq:Osrn} and \eqref{eq:Lbarn}
    \State
        Construct $\hat{\Osr}^{(k)}$ in \eqref{eq:Ohat} and $\hat{\Lbar}^{(k)}$ in \eqref{eq:Lhat}
    \If{$k>1$}
        \State Perform SSI-SVD $\hat{\Osr}^{(k)} = U^{(k)} \Sigma^{(k)} V^{(k)}$ by Algorithm \ref{algorithm-ssi}, with rank $\rank_{\max}$ and initial guess $U^{(k-1)}$
    \Else 
        \State Perform LMSVD $\hat{\Osr}^{(k)} = U^{(k)} \Sigma^{(k)} V^{(k)}$ with rank $\rank_{\max}$
    \EndIf
    \If{$\sigma_{\rank_{\max}} > r_{\rm reg} \sigma_{\max} $} 
        \State 
            $\rank_{\max} = (1+\epsilon) \rank_{\max}$
    \EndIf
    \State 
        Update $\OO^{(k)}$ and $\LL^{(k)}$ with \eqref{eq:Obar} and \eqref{eq:Lbar}.
    \State 
        Update $\param^{(k+1)}$ with \eqref{eq:updated_sr_1}
\EndFor
\end{algorithmic}
\hspace*{0.02in} 
{\bf Output:} $\param = \param^{(k_{\max})}$
\end{algorithm}

\subsection{Summary and comparisons with other SR variants}
\label{sec:comparison}

We now compare WSSR with other SR variants, highlighting the novelty of our method.
In particular we focus on comparisons with two state-of-the-art SR methods (described in Section \ref{sr:variants}): MinSR and SPRING.
The differences between WSSR and these methods primarily lie in the following three aspects.

The first difference lies in how historical information is averaged to construct the preconditioned gradient. 
The parameter update of the MinSR method depends solely on the current optimization step, discarding all historical information. 
The SPRING method reintroduces historical information by incorporating a momentum term and does not explicitly perform averaging of $\mS$. 
In contrast, our WSSR method averages both the $\mS$-matrix and the gradient within a unified and systematic framework.

The second difference resides in the regularization strategy of the approximate $\mS$-matrix. The MinSR method applies a diagonal shift to the $\mT$-matrix (defined in \eqref{eq:mT}). The SPRING method further improves numerical stability by adding a matrix with all elements set to $1/\sample$ to the $\mT$-matrix, in addition to the diagonal shift. On the other hand, our WSSR method dynamically truncates the singular values based on their decay and sets the remaining ones to a constant, allowing for efficient matrix inversion while ensuring good numerical stability.

The last difference consists in how the matrix inversion of the $\mS$-matrix is performed. Both the MinSR and SPRING methods compute the inverse of smaller matrices to achieve an equivalent result, which can be interpreted as an application of the Sherman-Morrison-Woodbury formula. The dominant cost $\mathcal{O}(\dof \sample^2)$ arises from forming $\mT_{\sample}(\param) := \Osr(\sse_{\sample};\param)^{\trans} \Osr(\sse_{\sample};\param)$, which exceeds the cost $\mathcal{O}(\sample^3)$ for the subsequent Cholesky factorization of $\mT_\sample$ \cite{goldshlager2024kaczmarz}.

In contrast, WSSR performs the inversion implicitly through an SSI–SVD applied to the averaged matrix $\hat{\Osr}^{(k)}$ without forming $\mT$ or $\mS$ matrix. After the SSI–SVD procedure produces the truncated singular vectors, the subsequent low-rank SR update only involves matrix–vector projections onto the resulting subspace and incurs a cost lower than the SSI–SVD itself. 

The SSI–SVD step requires 
$\mathcal{O}(m\dof \sample \rank)$ operations for the matrix–block multiplications and 
$\mathcal{O}(m\dof \rank^2)$ for orthonormalization, where 
$m$ denotes the number of subspace-iteration steps.
Since under our warm-start strategy 
$m$ remains small (often $3\sim5$) and 
$\rank \leq \min\{\dof, \sample\}$, the overall cost of the SSI–SVD procedure can be regarded as 
$\mathcal{O}(\dof \sample \rank)$.

By exploiting the similarity between the singular vectors of $\hat{\Osr}^{(k)}$ in consecutive VMC steps, warm-starting techniques enhances computational efficiency while maintaining accurate preconditioning.

\section{Numerical experiments}
\label{sec:numerics}

We apply our WSSR method for the ground state calculations of some atomic and molecular systems.
An efficient machine-learning architecture, the atomic cluster expansion (ACE), is employed to parameterize the many-electron wave functions \cite{drautz2022,zhou2024multilevel}. 
We provide the details of ACE parameterization in \ref{sec:ace}.
All results are given in atomic units (a.u.).
The simulations are performed via our {\sc Julia} implementations {\tt ACEpsi.jl} \cite{ACEpsi}. 
All simulations were performed on Digital Research Alliance of Canada. Each job was assigned a single NVIDIA H100 GPU.

Recall that $N$ denotes the number of electrons, $M$ the number nuclei and $Z$ the atomic number of a nucleus. 
We perform simulations for the beryllium atom with $N = 4$, $M = 1$, $Z = 4$, the oxygen atom with $N = 8$, $M = 1$, $Z = 8$, the nitrogen atom with $N = 10$, $M = 1$, $Z = 10$, the lithium-hydrogen (LiH) molecule with Li-H bond length of 3.015, and the lithium-lithium $\text{Li}_{\rm 2}$ molecule with a Li-Li bond length of 5.051.

We first test the choices of key hyperparameters: the relative cutoff $r_{\rm reg}$, the maximum rank $\rank_{\max}$, and the averaging weight $\delta$. 
We show in Fig. \ref{fig:comparison} 
the convergence behavior of the WSSR algorithm for the beryllium atom under different parameter choices.
For the regularization $r_{\rm reg}$, we find that smaller values yield faster convergence, accompanied with a relatively large rank in the adaptive selection scheme.
For the rank cutoff $\rank_{\max}$, the optimization becomes unstable and fails to converge reliably when $\rank_{\max} = 200$, and increasing $\rank_{\max}$ improves stability and yields lower energy values.
For the parameter averaging weight $\delta$, setting $\delta = 0.0$ leads to unstable optimization with  fluctuations, while $\delta = 0.5$ and $\delta = 0.95$ produce stable convergence behavior and slightly better convergence.
These results highlight the importance of incorporating historical information for stable and efficient optimization and indicate that excessively large $\rank_{\max}$ is unnecessary.

We compare the convergence of different SR methods in Fig. \ref{fig:time}. 
The hyperparameters used in these methods are presented in \ref{sec:parameters}. 
We observe that both the WSSR and SPRING methods achieve similar accuracy and convergence rates, whereas RSSR converges slightly more slowly.
We further compare in Table \ref{tab:time_comparison} the computational time required by the SPRING and WSSR methods across different atomic systems and basis sets.
For WSSR, increasing the rank $\rank$ slightly raises the computational cost, and the benefit becomes marginal beyond 
$\rank_{\max} = 800$, where the reduction in optimization time is minimal relative to the additional expense. 
These results demonstrate that WSSR offers superior computational efficiency while maintaining comparable accuracy for the case of wave function architectures based on the ACE framework. 

\begin{figure}[h]
\centering
\includegraphics[width = 6.6cm]{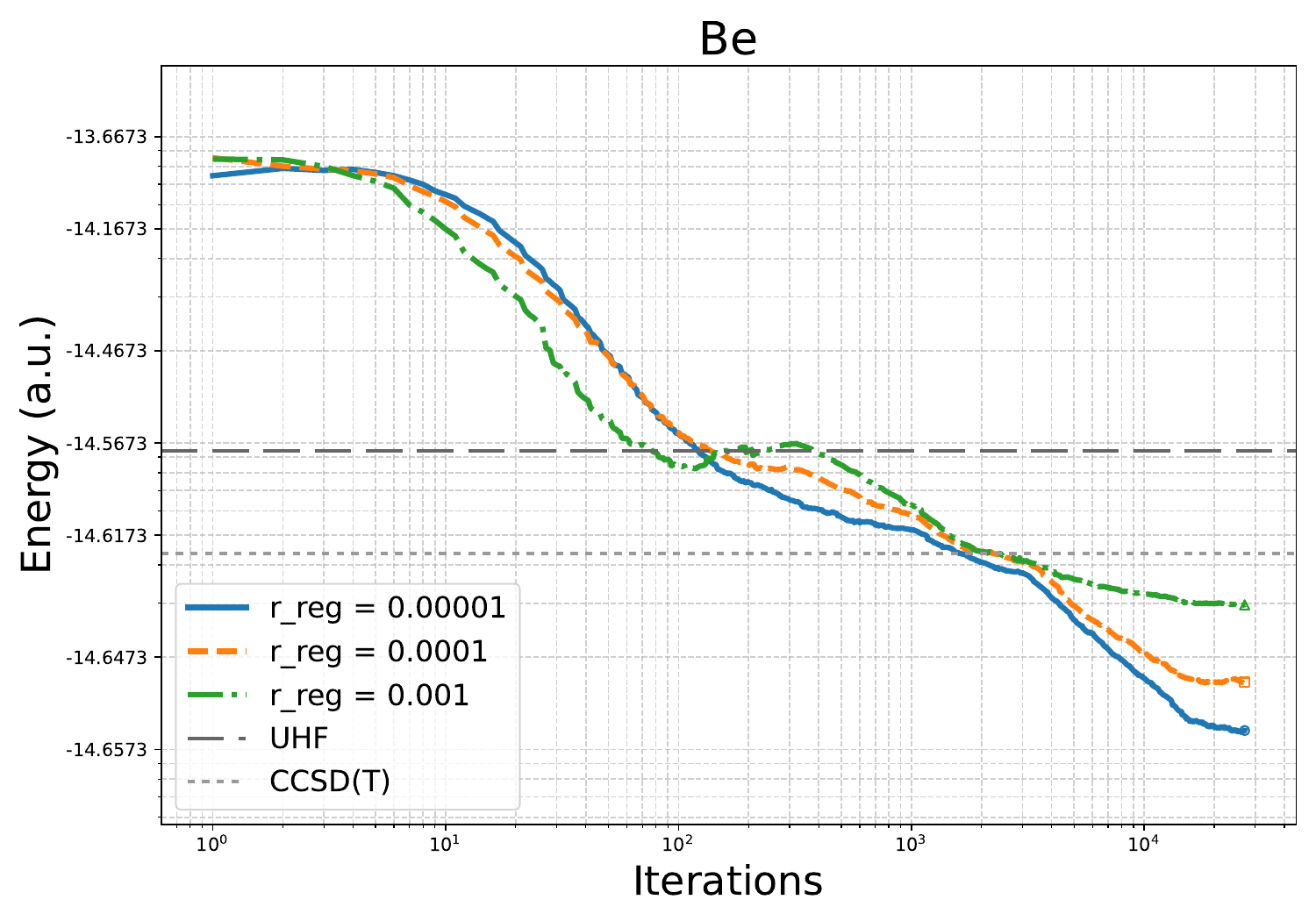}
\includegraphics[width = 6.6cm]{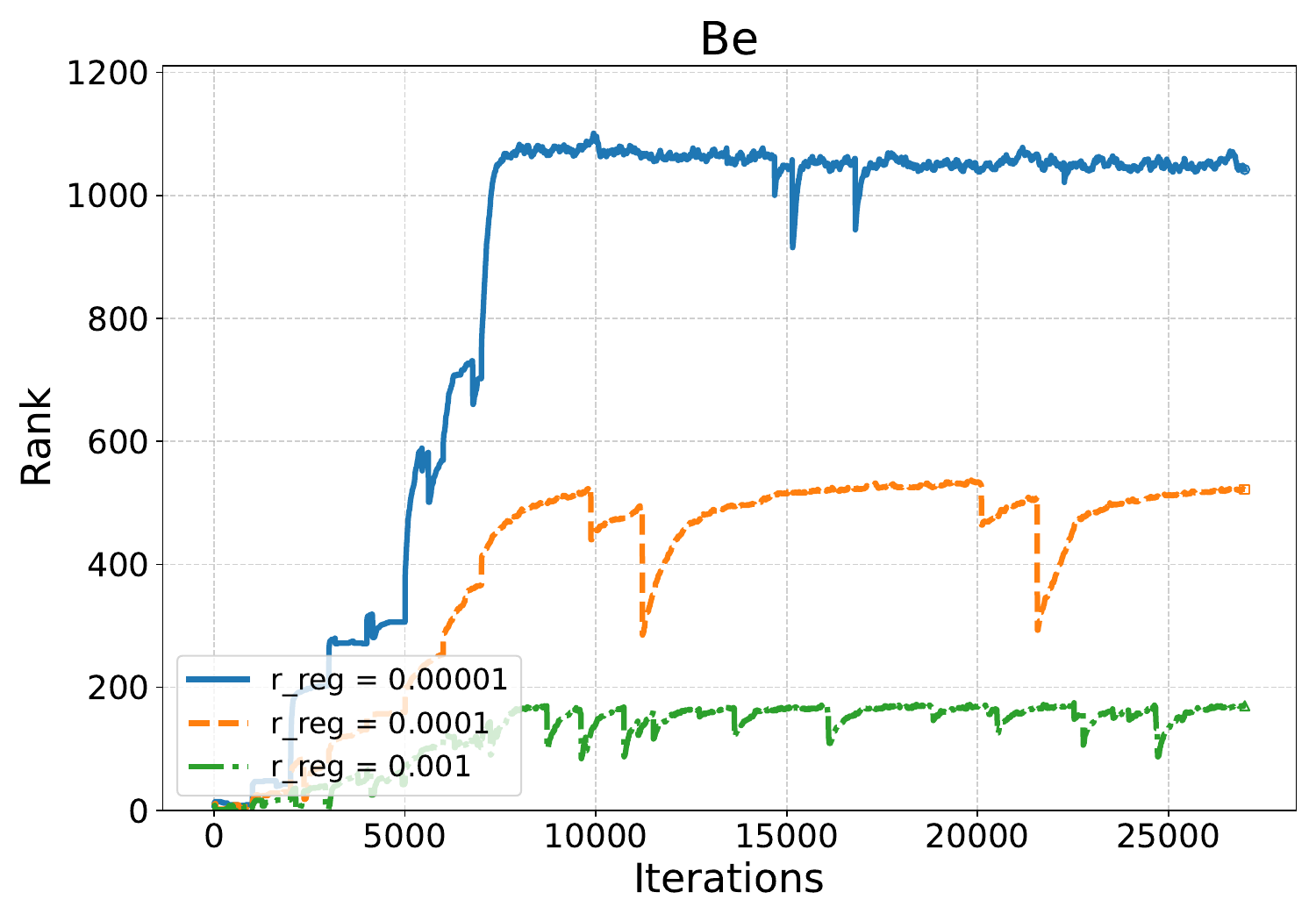}
\caption{Rank Selection for (beryllium atom).
Left: Relative cutoff $r_{\rm reg}$.
Right: Rank growth with iterations.
}
\label{fig:rankSelect}
\end{figure}

\begin{figure}[h]
\centering
\includegraphics[width = 6.6cm]{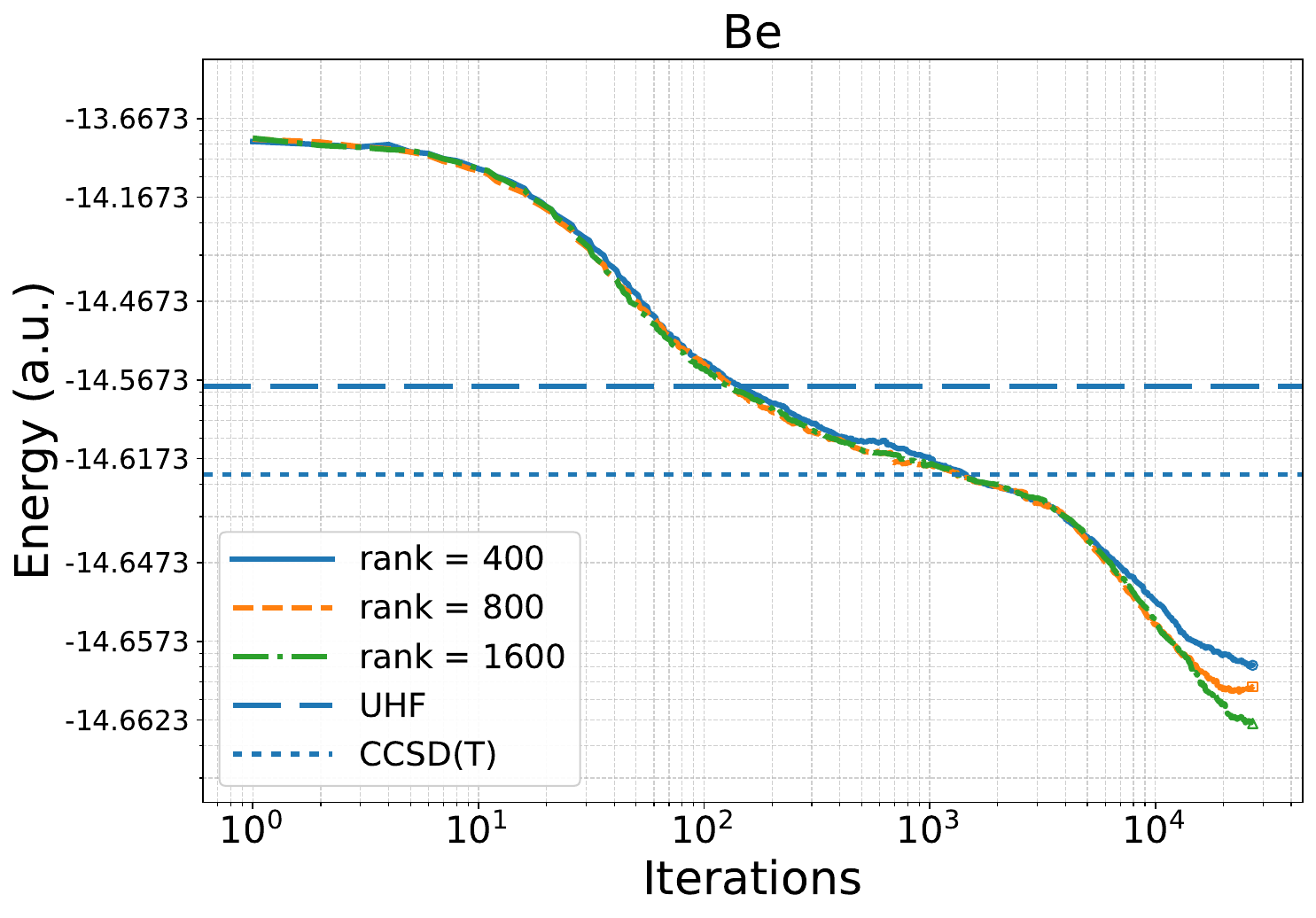}
\includegraphics[width = 6.6cm]{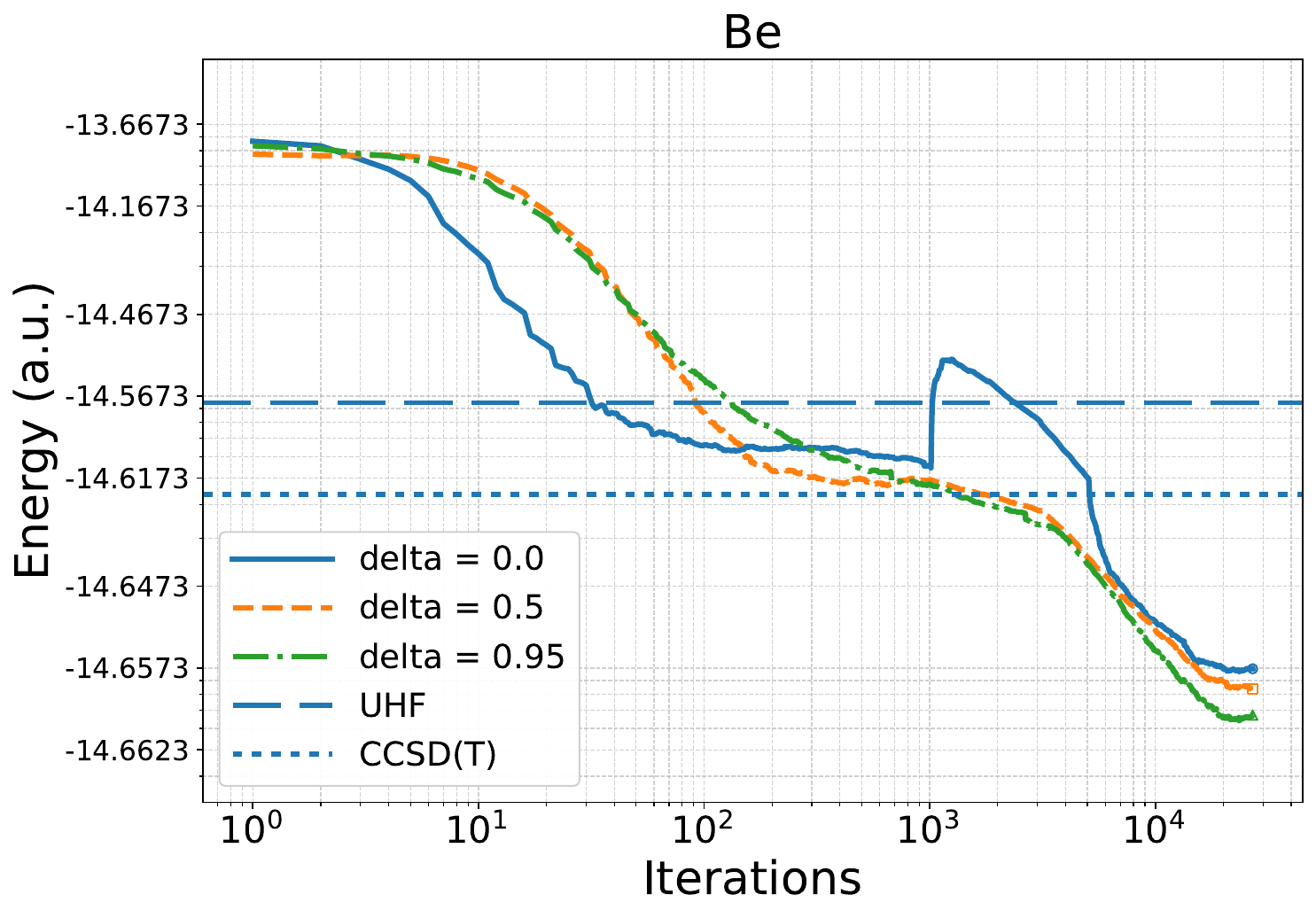}
\caption{Convergence with respect to different numerical parameters (beryllium atom).
Left: Maximum rank $\rank_{\max}$.
Right: Averaging weight $\delta$.}
\label{fig:comparison}
\end{figure}

\begin{figure}[h]
\centering
\includegraphics[width = 6.6cm]{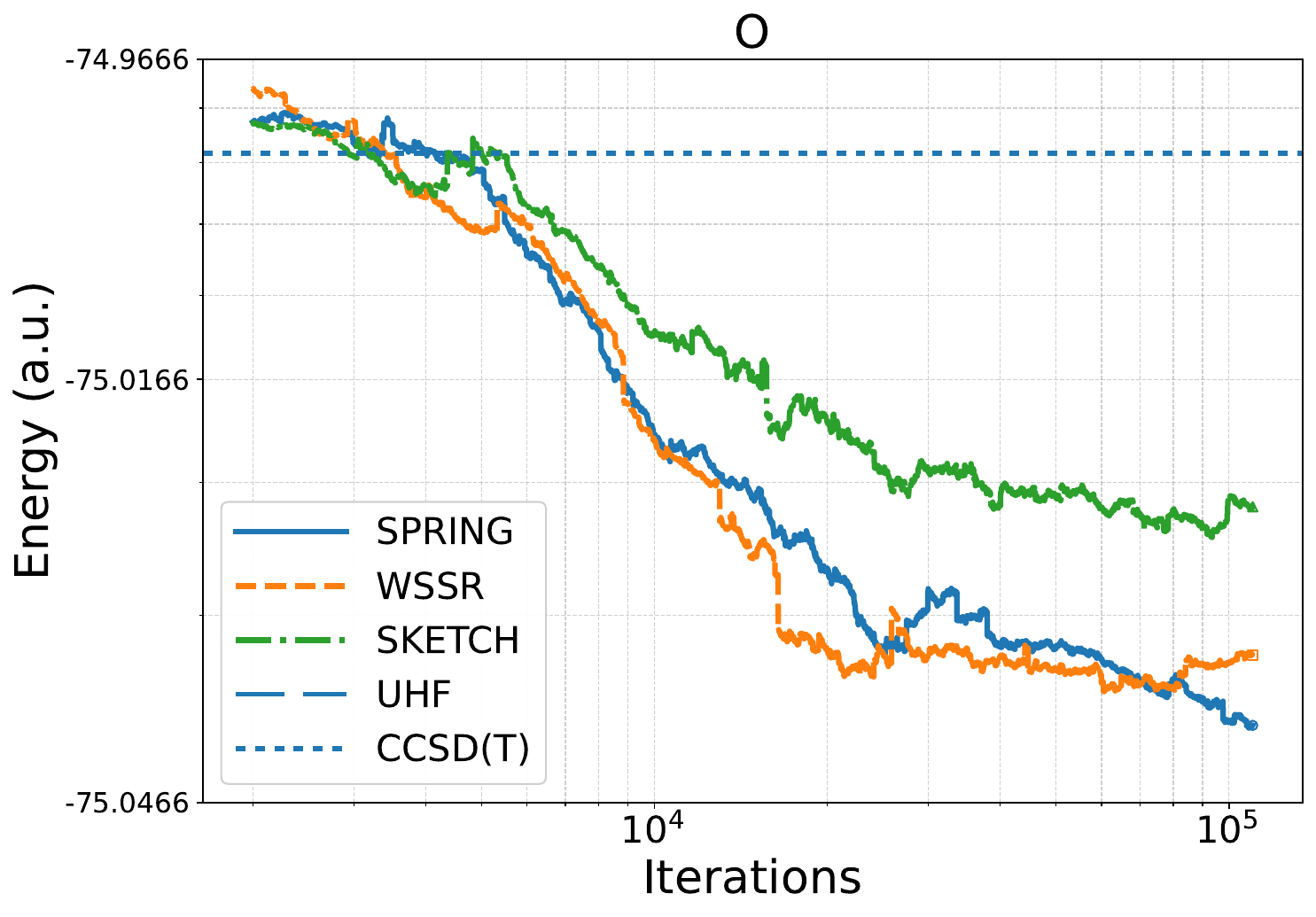}
\includegraphics[width = 6.6cm]{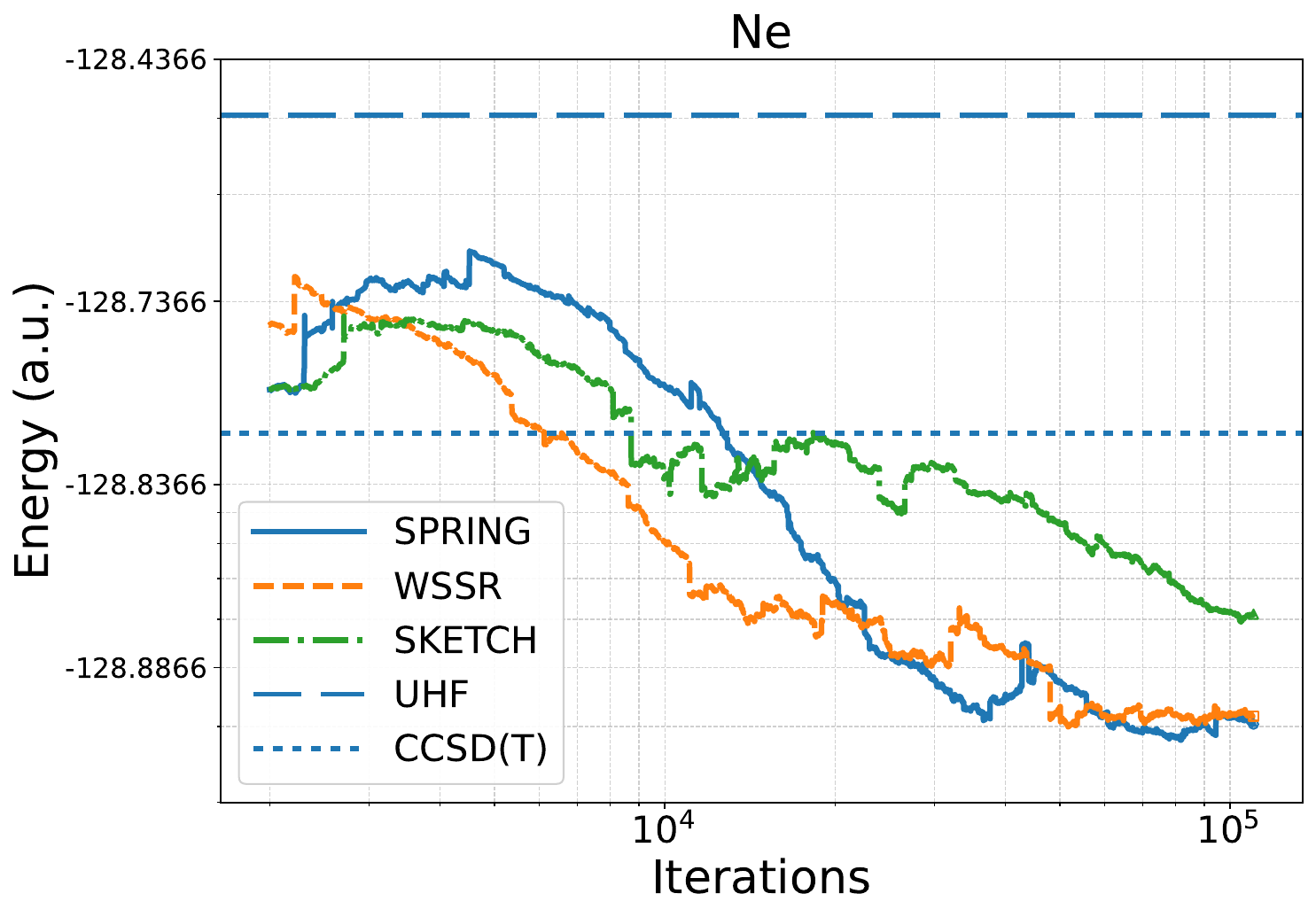}
\\ \vskip 0.2cm
\includegraphics[width = 6.6cm]{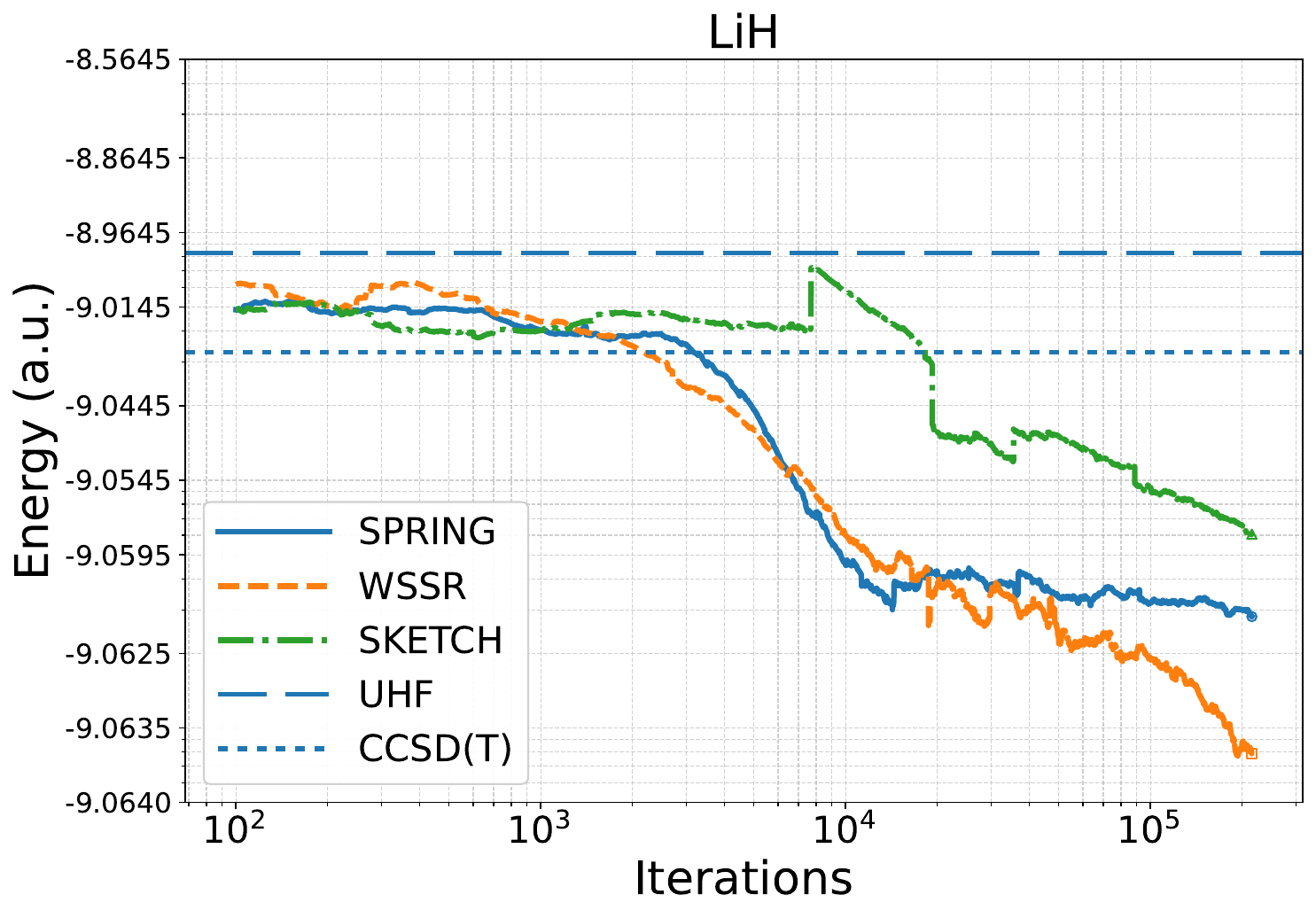}
\includegraphics[width = 6.6cm]{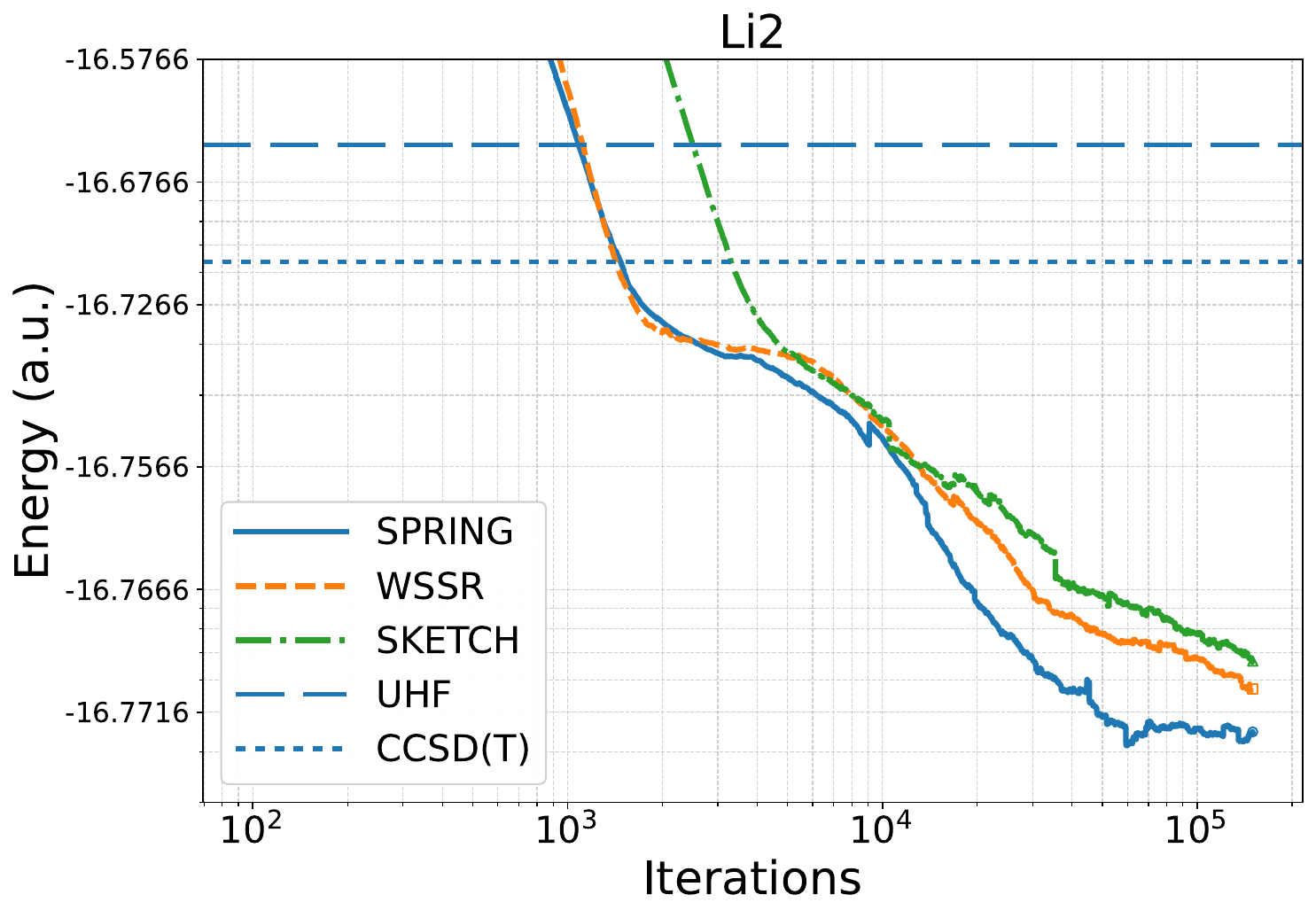}
\caption{Convergence comparison of different SR methods. 
First row (atoms): O; Ne.
Second row (molecule): LiH; $\text{Li}_{\rm 2}$. All UHF and CCSD(T) reference energies are computed using the cc-pVTZ basis set.}
\label{fig:time}
\end{figure}

\begin{table}[htbp!]
    \centering
    \renewcommand{\arraystretch}{1.0}
    \setlength{\tabcolsep}{5pt} 
    \small
    \begin{tabular}{ll|c|cccc}
       \toprule
        System & Basis & SPRING &  \multicolumn{4}{c}{WSSR} \\
        \cmidrule(lr){4-7}
        & & & $~~200~~$ & $~~400~~$ & $~~800~~$ & $~~1600~~$ \\
        \midrule
        \multirow{2}{*}{Be} & DZ &  3.98 & 0.88 & 1.25  &    1.78  & 3.39  \\
        & TZ & 7.20 & 1.74 & 3.02 & 5.37  & 10.39     \\
        \midrule
        \multirow{2}{*}{O}  & DZ & 3.50 &  1.08 & 1.56 & 2.30  &   4.59   \\
                            & TZ & 8.16 &  1.80 & 3.26 &  6.11  & 14.65 \\
        \midrule
        \multirow{2}{*}{Ne} & DZ & 3.64 &  0.92 & 1.56 & 2.57 & 5.40   \\
                            & TZ & 7.16 &  1.70 & 3.20 & 5.68 & 13.46 \\
        \bottomrule
    \end{tabular}
    \caption{Comparison of computational time between SPRING and WSSR for different atomic systems and basis sets. The WSSR method is evaluated for different values of $\rank_{\max}$. The baseline unit is defined as the time required for VMC calculations, excluding the optimization steps (sampling, gradient computation, and energy calculation). For each iteration, SVD of rank $\rank_{\max}$ is computed, following with the application of truncation criterion \cref{eq:ruler}, see Section~\ref{sec:rank} for details. }
    \label{tab:time_comparison}
\end{table}

\section{Conclusions}
\label{sec:conclusions}
In this work, we develop the WSSR optimizer for VMC ground state calculations of many-electron systems, which incorporates a warm-started SVD algorithm to accelerate the matrix inversions in the stochastic reconfiguration method.
By leveraging optimization history and adaptively increasing the rank, this method significantly reduces computational costs.
Our numerical experiments show that the WSSR method outperforms alternative SR methods in terms of convergence rate and computational time.

\section*{Acknowledgements}
The work of Dexuan Zhou was supported by National Natural Science Foundation of China (124B2020). The work of Huajie Chen was supported by the National Natural Science Foundation of China (12371431). The work of Xin Liu was supported in part by the National Natural Science Foundation of China (12125108, 12288201), and RGC grant JLFS/P-501/24 for the CAS AMSS–PolyU Joint Laboratory in Applied Mathematics. Christoph Ortner was supported by the Natural Sciences and Engineering Research Council of Canada through an NSERC Discovery grant. This research was supported in part through computational resources and services provided by the Digital Research Alliance of Canada (alliancecan.ca) and by Advanced Research Computing at the University of British Columbia. Conflict of interest disclosure: Christoph Ortner is a partner in Symmetric Group LLP which licenses force fields commercially. 

\appendix

\section{SVD with Randomized Sketching}
\label{sec:SVD:random}

In this section, we give a brief overview of the randomized sketching scheme for approximating the singular value decomposition (SVD) of a $\dof \times \sample$ matrix $\hat{\Osr}^{(k)}$ with rank $\rank$, following \cite{halko2011finding}.

Given a target rank $\rank$ and an oversampling parameter $p$, our first goal is to construct a matrix $Q$ with $\rank + p$ orthonormal columns such that
\begin{eqnarray*}
    \|\hat{\Osr}^{(k)} - QQ^{\trans} \hat{\Osr}^{(k)} \| \approx \min_{\text{rank}(X)\leq \rank}\|\hat{\Osr}^{(k)} - X\|. 
\end{eqnarray*}
The matrix $Q$ spans a subspace that captures most of the action of $\hat{\Osr}^{(k)}$. When $p = 0$, the columns of $Q$ are exactly the dominant left singular vectors of $\hat{\Osr}^{(k)}$. For $p > 0$, the additional flexibility improves computational efficiency. 

To construct the matrix $Q$, we first generate $\rank + p$ random samples as 
\begin{eqnarray*}
\Omega = (\omega^{(1)}, \omega^{(2)}, \cdots, \omega^{(\rank + p)}) \in \R^{\dof \times (\rank+p)}. 
\end{eqnarray*}
These random vectors are likely to form a linearly independent set, and no linear combination falls within the null space of $\hat{\Osr}^{(k)}$. 
Consequently, the columns of $Y = \hat{\Osr}^{(k)} \Omega$ are expected to be linearly independent, spanning the range of $\hat{\Osr}^{(k)}$. 
The random matrix $\Omega$ can be chosen in various ways, such as a Gaussian random matrix. However, using a structured random matrix, such as a subsampled random Fourier transform (SRFT) defined in Section 4.6 of \cite{halko2011finding}, helps reduce the computational cost of the product $Y = \hat{\Osr}^{(k)} \Omega$ to $O(\dof \sample \log(\rank))$.

With fixed-rank approximation, the rank-$\rank$ approximation of $\hat{\Osr}^{(k)}$ is obtained by performing SVD on the reduced matrix $B = Q^{\trans} \hat{\Osr}^{(k)} \in \R^{\rank \times \sample}$:
\begin{align}
\label{eq:randsvd}
    \hat{\Osr}^{(k)} \approx U^{(k)} \Sigma^{(k)} V^{(k)}, \quad {\rm with}~U^{(k)} = QU. 
\end{align} 
where
\begin{align*}
    B = U\Sigma^{(k)} V^{(k)}. 
\end{align*}
This procedure is computationally efficient, particularly when $Q$ has relatively few columns (i.e., when $\rank + p$ is small). In such cases, the reduced matrix $B$ is easily constructed, and its SVD can be computed quickly.

In this algorithm, the appropriate of $Q$ by a structured random matrix takes $O(\dof \sample \log(\rank) + \dof\rank^2)$; the SVD calculation of $B$ with row-extraction technique \cite{halko2011finding} costs $O((\dof+\sample)\rank^2)$. 
Compared with deterministic techniques of SVD calculation requiring $O(\dof\sample\rank)$, the randomized algorithm can be several times faster and is well suited for parallel implementation \cite{halko2011finding}.

\section{Wavefunction Architecture with ACE}
\label{sec:ace}

In this section, we provide a brief overview of the parameterization of wave functions using Atomic Cluster Expansion (ACE). For details, refer to \cite{zhou2024multilevel}.

To construct ACE wave function parameterization, we first construct one-body basis functions. Let $\basis_{\ind}:\R^3\times\Z_2\rightarrow\R$ denote one-body basis functions, where $\ind$ denotes the index of the basis. 
In this work, we focus on the atomic orbital \cite{helgaker2014molecular}. 

With those one-body basis functions set, we define an $N$-variable symmetric function with index $\ind$ as
\begin{equation*}
   A_\ind(\xx_1,\cdots,\xx_N) := \sum_{i=1}^N \basis_\nu(\xx_i). 
\end{equation*}
The ACE basis functions take the form of a product of these functions
\begin{equation*}
    \ace_{\aceind}(\pmb{x}) := \prod_{t=1}^{N} A_{\ind_t}(\pmb{x}),
\end{equation*}
where the index set $\aceind$ belongs to
\begin{equation*}
\aceI_N := \Big\{\aceind = (\ind_1, \cdots, \ind_N) \in (\mathcal{K}\times \Z_2)^N : \ind_1 \leq \ind_2 \leq \cdots \leq \ind_N \Big\}.
\end{equation*}
To further reduce complexity, we truncate the expansion by considering only clusters of up to $\aceB$ particles, which leads to the modified basis functions, 
\begin{equation*}
\ace_{\ind_1, \dots, \ind_\aceB}(x_1,\cdots,x_N) := \prod_{t=1}^{\aceB} A_{\ind_t}(x_1,\cdots,x_N).
\end{equation*}
$N$-particle symmetric function $\Phi(\xx)$ can then be approximated by ACE model as 
\begin{equation*}
\Phi(\xx) \approx \sum_{\aceind\in \aceI_{\aceB}} c_{\aceind} ~ \ace_{\aceind}(\xx),
\end{equation*}
with coefficients $\param:=\{c_{\aceind}\}_{\aceind\in \aceI_\aceB}$. 

To construct antisymmetric wave functions, 
we further denote $\widetilde{\ace}_{\aceind}$ as an ACE basis function in which the particle $\xx_i$ is treated as the ``highlighted" particle. 
Specifically, for an index tuple 
$\aceind = (\nu_0,\nu_1,\ldots,\nu_{\aceB-1})$, we define
\begin{equation*}
    \widetilde{\ace}_{\aceind}(\xx_i;\xx_{\neq i})
    :=
    \basis_{\nu_0}(\xx_i)
    \prod_{t=1}^{\aceB-1} 
        \left(
            \sum_{j\neq i} \basis_{\nu_t}(\xx_j)
        \right),
\end{equation*}
so that the first index $\nu_0$ acts on the highlighted particle $\xx_i$, 
while each $\nu_t$ ($t\ge 1$) corresponds to a symmetric sum of one-body basis functions 
over all other particles $\xx_j$ with $j\neq i$. 

With these partially symmetric basis functions $\widetilde{\ace}_{\aceind}$, we define backflow determinant as 
\begin{equation*}
\wfbf(\xx) = \det \begin{pmatrix}
\varphi_1(\xx_1;\xx_{\neq 1}) & \cdots & \varphi_{N}(\xx_1;\xx_{\neq 1}) \\
\vdots & \ddots & \vdots \\
\varphi_1(\xx_{N};\xx_{\neq N})  & \cdots & \varphi_{N}(\xx_{N};\xx_{\neq N})
\end{pmatrix},
\end{equation*}
where $\varphi_k$ takes the form 
\begin{equation*}
\varphi_k\big(\xx_{i};\xx_{\neq i}\big) 
\approx \sum_{\aceind\in \widetilde{\aceI}_{\aceB}} \tilde{c}^k_{\aceind} ~ \widetilde{\ace}_{\aceind}\big(\xx_i; \xx_{\neq i}\big).   
\end{equation*}
Finally, the full wave function used in this work takes the form
\[
\Psi(\xx)
= 
\wfbf(\xx)\,
\exp\big(\gamma(\xx)\big),
\]
where $\gamma(\xx)$ is a Jastrow factor that enforces the electronic cusp
conditions,
\[
\gamma(\xx)
:= 
\sum_{i<j}
\frac{-c_{ij}}{1 + \lvert \rr_i - \rr_j\rvert},
\qquad
c_{ij}
=
\begin{cases}
\frac{1}{2}, & \text{opposite spins},\\[2pt]
\frac{1}{4}, & \text{same spin},
\end{cases}.
\]

\section{Hyperparameters}
\label{sec:parameters}

We list all hyperparameters in \cref{tab:training_settings}. Specifically, we employ a decaying learning rate, defined as $\eta_k=\alpha/(1+k/\beta)$ with fixed parameters $\alpha= 0.015$ and $\beta= 1000.0$. 
The maximum number of training steps is set to $100,000$ to ensure robust convergence across all systems. 
For SPRING, we adopt a Tikhonov regularization parameter of $\epsilon = 0.001$, and a regularization parameter $\mu = 0.99$. 
In our numerical experiments, we typically set the maximum number of SSI iterations to be $3$ perform orthonormalization every step. 
At the first step for each level of the VMC optimization, we use the LMSVD method \cite{liu2013limited} to compute a more accurate SVD, ensuring a better initial guess for the subsequent steps.
Finally, we smooth the energy curves for each method by averaging over a sliding window of ten thousand iterations. 
In all numerical experiments, we employed a multilevel wave function structure to ensure a robust initialization. 

\begin{table}[htbp!]
    \centering
    \begin{tabular}{|c|c|}
        \hline
        \textbf{Hyperparameters} & \textbf{Value} \\ \hline
        Standard deviations for local energy clipping & 5 \\ \hline
        Number of walkers & 2048 \\ \hline
        MCMC burn-in steps & 1000 \\ \hline
        MCMC steps between updates & 10 \\ \hline
        MCMC Sampling distribution & Gaussian distribution \\ \hline
        SSI Maximum iterations & 3 \\ \hline  
    \end{tabular}
    \label{tab:training_settings}
\end{table}

\small
\bibliographystyle{elsarticle-harv} 
\bibliography{references}

\end{document}